\documentclass[11pt,english,a4paper,floatfix]{article}
\usepackage{jcappub}
\usepackage{multirow}

\pdfoutput=1
\usepackage{bm}
\usepackage{amsfonts}

\newcommand{\CAMB}{\textsc{camb}}
\newcommand{\logg}{\log_{10}\left[G_\text{eff}\text{MeV}^2\right]}
\newcommand{\mnu}{\sum m_\nu}
\newcommand{\nnu}{N_\text{eff}}
\newcommand{\MIn}{\text{MI}\nu}
\newcommand{\SIn}{\text{SI}\nu}

\begin{document}


\title{Updated constraints on massive neutrino self-interactions from cosmology in light of the $H_0$ tension}

\author[a]{Shouvik Roy Choudhury,}
\author[b]{Steen Hannestad,}
\author[b]{Thomas Tram}

\affiliation[a]{Department of Physics, Indian Institute of Technology Bombay,
Main Gate Road, Powai, Mumbai 400076, India}
\affiliation[b]{Department of Physics and Astronomy, Aarhus University,
 DK-8000 Aarhus C, Denmark}

\emailAdd{shouvikroychoudhury@gmail.com}
\emailAdd{sth@phys.au.dk}
\emailAdd{thomas.tram@phys.au.dk}

\abstract{
We have updated the constraints on flavour universal neutrino self-interactions mediated by a heavy scalar, in the effective 4-fermion interaction limit. We use the relaxation time approximation to modify the collisional neutrino Boltzmann equations, which is known to be very accurate for this particular scenario. Based on the latest CMB data from the Planck 2018 data release as well as auxiliary data we confirm the presence of a region in parameter space with relatively strong self-interactions which provides a better than naively expected fit. However, we also find that the most recent data, in particular high-$\ell$ polarisation data from the Planck 2018 release, disfavours this solution even though it cannot yet be excluded. Our analysis takes into account finite neutrino masses (parameterised in terms of $\sum m_{\nu}$) and allows for a varying neutrino energy density (parameterised in terms of $N_{\rm eff}$), and we find that in all cases the neutrino mass bound inferred from cosmological data is robust against the presence of neutrino self-interactions. Finally, we also find that the strong neutrino self-interactions do not lead to a high value of $H_0$ being preferred, i.e.\ this model is not a viable solution to the current $H_0$ discrepancy.
}

\maketitle


\section{Introduction}

Neutrinos are massless in the standard model of particle physics, but terrestrial neutrino oscillation experiments~\cite{Fukuda:1998mi,Ahmad:2002jz} have long established the presence of at least 3 non-degenerate neutrino mass eigenstates (with at least two of the masses being non-zero) which are quantum super-positions of their flavour eigenstates. While neutrino oscillations data is sensitive to the neutrino mass-squared splittings, cosmological data is sensitive mainly to the sum of neutrino masses parameter, typically denoted by $\sum m_{\nu}$. At present, the tightest bounds on $\sum m_{\nu}$ is around $\sum m_{\nu} \lesssim 0.12$ eV (95\% C.L.)~\cite{Aghanim:2018eyx,RoyChoudhury:2019hls,Choudhury:2018byy,Alam:2020sor,Vagnozzi:2017ovm} under the assumption of a $\Lambda\textrm{CDM}+\sum m_{\nu}$ cosmology with 3 degenerate neutrino masses.

There are several models that have been proposed to explain the generation of neutrino masses, and one such interesting direction is when we consider the neutrinos to be Majorana particles, and the $U(1)_{B-L}$~\cite{Gelmini:1980re,Choi:1991aa,Acker:1992eh,Chikashige:1980ui,Georgi:1981pg} symmetry is spontaneously broken. The symmetry breaking gives rise to a new Goldstone boson, the majoron, denoted by $\phi$. It couples to the neutrinos via the Yukawa interaction~\cite{Oldengott:2017fhy,Oldengott:2014qra},

\begin{equation}
\mathcal{L}_{\rm int} = g_{ij} \bar{\nu_i} \nu_j \phi + h_{ij}  \bar{\nu_i} \gamma_5 \nu_j \phi,
\end{equation} \label{eq1}
where $g_{ij}$ and $h_{ij}$ are, respectively, the scalar and pseudo-scalar coupling matrices, $\nu_i$ is a left-handed neutrino Majorana spinor, and the indices $i,j$ are used to label the individual neutrino mass eigenstates. While we have motivated such an interaction via the majoron-like model of neutrino mass generation, we emphasise here that in general this kind of interaction is not limited to such models. For instance, $\phi$ can be linked to the dark sector~\cite{Barenboim:2019tux}.

In this paper we consider the special case where $g_{ij} = g \delta_{ij}$ and $h_{ij} = 0$, where $\delta_{ij}$ is the Kronecker delta. Hence we see that the matrix $g_{ij}$ has the same form in both flavour and mass basis. Such simplification might be unrealistic for real particle physics models, but it gives us a simple way to test the sensitivity of the cosmological data to such neutrino-majoron couplings~\cite{Kreisch:2019yzn}. We also restrict ourselves to the situation where the scalar mass, $m_{\phi}$ is much larger than the energies of neutrinos during the formation of the Cosmic Microwave Background (CMB) radiation
\footnote{When $m_\phi \sim T$ or smaller the phenomenology of the model changes significantly: The system undergoes recoupling instead of decoupling, and a new population of $\phi$ particles can be built up from neutrino pair annihilation. We refer the reader to e.g.\ Refs.\ \cite{Archidiacono:2013dua,Forastieri:2019cuf,Archidiacono:2020yey} for a more detailed discussion.}.
In that case the interaction Lagrangian in equation~(\ref{eq1}) effectively becomes a 4-fermion interaction, i.e. we can treat it as a $\nu \nu \rightarrow \nu \nu$ self-interaction with a self-interaction rate per particle $\Gamma \sim g^4 T_{\nu}^5/m_{\phi}^2 = G_{\rm eff}^2 T_{\nu}^5$, where $G_{\rm eff} = g^2/m_{\phi}^2$ is the effective self-coupling~\cite{Oldengott:2017fhy}. Any initial population of $\phi$ in the early universe can be considered to have completely decayed by the time of the CMB formation epoch. This can be ensured with a typical $m_{\phi} \gtrsim$ keV~\cite{Blinov:2019gcj}. Again, non-standard neutrino interactions in the limit where the mediator is heavy are more general than what we have considered here (see~\cite{Ohlsson:2012kf}). Considering a heavy vector boson (as in~\cite{Archidiacono:2013dua}) will lead to a similar four-fermion interaction and all the cosmological implications drawn in this paper will apply to such interactions as well.  For such non-standard interactions with a heavy mediator, while the neutrinos as usual decouple from the primordial plasma at the decoupling temperature $T \sim 1$ MeV (determined via the weak interaction coupling strength), they continue to scatter with each other after decoupling (instead of free-streaming when there is no such self-interaction) with the interaction rate $\Gamma$, assuming $G_{\rm eff} > G_{\rm W}$, where $G_{\rm W} \simeq 1.166 \times 10^{-11} \textrm{MeV}^{-2}$ is the standard Fermi constant. The self-scattering will also eventually die out when $\Gamma$ falls below the Hubble rate. Thus increasing $G_{\rm eff}$ would further delay the neutrino free-streaming.

Cosmological constraints on $G_{\rm eff}$ have been previously studied in~\cite{Oldengott:2017fhy,Barenboim:2019tux,Cyr-Racine:2013jua,Archidiacono:2013dua,Lancaster:2017ksf,Park:2019ibn,Kreisch:2019yzn}. In the $\Lambda\rm CDM + \textrm{log}_{10} \left[G_{\rm eff} \textrm{MeV}^2\right]$ (i.e. $\Lambda\rm CDM$ model augmented with self-interacting but massless neutrinos), when tested with older versions of the Planck likelihoods, one finds a bimodal posterior for the $\textrm{log}_{10} \left[G_{\rm eff} \textrm{MeV}^2\right]$ parameter. As seen in~\cite{Oldengott:2017fhy} the Moderately Interacting Mode (hereafter MI$\nu$) typically corresponds to $\textrm{log}_{10} \left[G_{\rm eff} \textrm{MeV}^2\right] < -2.8$ (95\% C.L.) with CMB data from Planck 2015 temperature, polarisation measurements and Baryon Acoustic Oscillations (BAO), whereas the Strongly Interacting Mode (hereafter SI$\nu$) corresponds to $\textrm{log}_{10} (G_{\rm eff} \textrm{MeV}^2) = -1.5^{+0.3}_{-0.4}$ (68\% C.L.), and the posterior falls to almost zero in between the two regions. For the SI$\nu$, $G_{\rm eff} \simeq 3 \times 10^9 G_{W}$. Such strong interactions are allowed in the CMB data SI$\nu$ mainly through a degeneracy present among $G_{\rm eff}$, the angular size of the sound horizon at the last scattering $\theta_s$ and the scalar spectral index $n_s$. This degeneracy leads to bimodal posterior distributions with distinct modes in these three parameters as well. Strong interactions due to large $G_{\rm eff}$ pertain to a lack of anisotropic stress in the neutrino sector, the effect of which on the CMB power spectra can be compensated partially by increasing $\theta_s$.  At the same time, increasing $G_{\rm eff}$ causes a gradual increase in the power in small scales of the CMB power spectrum which can be partially compensated by a smaller $n_s$~\cite{Oldengott:2017fhy}.

In this paper, however, we are interested in a cosmological model that has been extended further: $\Lambda\rm CDM + \textrm{log}_{10} \left[G_{\rm eff} \textrm{MeV}^2\right] + N_{\rm eff}$ + $\sum m_{\nu}$, where, for our purposes, $N_{\rm eff}$ is the effective number of neutrino species (in general, it constitutes any relativistic species other than photons, in the early universe). $G_{\rm eff}$ is degenerate with both $N_{\rm eff}$ and $\sum m_{\nu}$ in the CMB power spectra, and~\cite{Kreisch:2019yzn} have shown that for the SI$\nu$ case, this extended model allows for high values of $N_{\rm eff}$ and $H_0$ that can provide a pre-recombination solution to the well-known $H_0$ tension (see~\cite{Knox:2019rjx} for a review), when tested with Planck 2015 high-$l$ and low-$l$ temperature data, combined with Planck 2015 lensing and BAO. Thus, varying both $N_{\rm eff}$ and $\sum m_{\nu}$ in the model was essential for the reconciliation of the $H_0$-tension. However, the analyses in~\cite{Kreisch:2019yzn} are done with the ``lite'' version of the high-$l$ Planck likelihoods where all but one of the nuisance parameters are marginalised over. Also, as seen from the tables VI and VII of~\cite{Kreisch:2019yzn}, when the lite version of Planck 2015 polarization likelihood is used with temperature, the $H_0$ values remain low even in the case of SI$\nu$ model. Thus, while very promising, the efficacy of this model in solving the Hubble tension remained doubtful. Neutrino self-interactions are also strongly constrained from particle physics experiments, with the exception of flavour specific interaction among the $\tau$-neutrinos \cite{Blinov:2019gcj,Brdar:2020nbj,Lyu:2020lps,Berbig:2020wve}. 

Currently, the $H_0$ value inferred from the Cepheid calibrated Type Ia Supernovae (SNe~Ia) $H_0=74.03 \pm 1.42$ km/s/Mpc (68\% C.L.)~\cite{Riess:2019cxk} (hereafter R19) by the SH0ES team is in 4.4$\sigma$ tension with the measurement of $H_0 = 67.27 \pm 0.60$ km/s/Mpc (68\% C.L.)~\cite{Aghanim:2018eyx} in the $\Lambda\rm CDM$ model with Planck 2018 CMB temperature and polarization power spectra. It is very plausible that the $H_0$ discrepancy is due to some physics beyond the basic $\Lambda$CDM model.
On the other hand, SNe~Ia calibrated with the Tip of the Red Giant Branch (TRGB) lead to a value of $H_0 = 69.8 \pm 0.8 (\text{stat.}) \pm 1.7 (\text{sys.})$ km/s/Mpc (68\% C.L.), which is within the 2$\sigma$ of the above two discrepant values.

With the release of the final Planck 2018 likelihoods~\cite{Aghanim:2019ame}, we find it timely and relevant to reassess the neutrino self-interaction situation from the cosmological perspective, and its association with the $H_0$-tension. Thus, in this work, we have updated the results on the neutrino self interactions in the  $\Lambda\rm CDM + \textrm{log}_{10} \left[G_{\rm eff} \textrm{MeV}^2\right] + N_{\rm eff}$ + $\sum m_{\nu}$ model using the full Planck 2018 temperature and polarization likelihoods, and with additional data. We find that, while the posterior for $\textrm{log}_{10} \left[G_{\rm eff} \textrm{MeV}^2\right]$ is still bimodal, even for the strongly interacting mode (SI$\nu$) high $N_{\rm eff}$ and high $H_0$ values that can solve the Hubble tension are \textit{not} preferred by the data. However, without the inclusion of the Planck 2018 high-$l$ TE, EE spectra, the SI$\nu$ mode does still allow $N_{\rm eff}$ values $> 4$ and large $H_0$ values encompassing the R19 bound (while the MI$\nu$ mode does not). We emphasize here that a theoretical model should only be considered a solution to the $H_0$-tension, if it is able to reconcile the  $H_0$-value obtained from the analysis of the \textit{full} Planck data combined with BAO, SNe, and LSS probes, with that from the R19 measurement. While these above results remain similar with the further inclusion of Planck 2018 CMB lensing, BAO and RSD measurements from SDSS-III BOSS DR12~\cite{Alam:2016hwk}, BAO measurements from MGS \cite{Ross:2014qpa} and 6dFGS \cite{Beutler:2011hx}, and SNe~Ia luminosity distance measurements from the Pantheon sample~\cite{Scolnic:2017caz},  with the latest full Planck likelihoods, the strongly interacting neutrinos do not seem to fulfill this criterion for a solution to the $H_0$-tension. 

Note that recently, three other papers~\cite{Brinckmann:2020bcn,Das:2020xke,Mazumdar:2020ibx} also tested the non-standard neutrino interactions against the latest Planck likelihoods. Qualitatively, our results agree quite well with \cite{Brinckmann:2020bcn}. In ref.~\cite{Das:2020xke} flavour universality of the self-interactions was not assumed. Instead, the self-interaction is taken to pertain to either 1,2, or 3 species which is more general than what is assumed in the present work. However the cosmological model considered in~\cite{Das:2020xke} is $\Lambda\rm CDM + \textrm{log}_{10} \left[G_{\rm eff} \textrm{MeV}^2\right]$ with massless neutrinos and $N_{\rm eff}$ fixed to 3.046, i.e. the parameters $N_{\rm eff}$ and $\sum m_{\nu}$ are not varied in their cosmological parameter analysis. Overall, our results are consistent with~\cite{Das:2020xke} in the limits where they can be compared (i.e.\ the scenario with flavour universal couplings, called \textbf{3c+0f}). However, \cite{Mazumdar:2020ibx} does a separate analysis with IceCube \cite{Aartsen:2020aqd} data which shows the SI$\nu$ region is disfavoured for both flavour specific and universal cases.

The rest of the paper is structured as follows. In section~\ref{section:2} we revisit the modification to the cosmological perturbation equations due to the neutrino interaction model, the cosmological model parametrisation and priors and the analysis method adopted, as well as the cosmological datasets used in this paper. In section~\ref{section:3} we present the results of the analyses and in section~\ref{section:4} we conclude.

\section{Methodology}
\label{section:2}
We will use the scalar mediated interaction in the effective 4-fermion interaction limit, as discussed in the previous section as our model for the neutrino-neutrino interaction. While this implies quite specific values for the various interaction terms we again note that our results will be generally applicable to other type of non-standard interactions with heavy mediators.

Using the scalar interaction we have modified the cosmological perturbation equations and implemented them in the \CAMB{} \cite{Lewis:1999bs} code.
We will assume throughout that the interaction has no effect on the background evolution. For the case of a heavy mediator this should be an excellent approximation since it can be safely assumed that any pre-existing population of the heavy mediator has decayed away before the CMB epoch, and that pair-annihilation/production can be neglected because of the large rest mass of the mediator particle.

\subsection{Cosmological perturbation equations}

Following previous work on the subject we will use the relaxation time approximation (RTA) first introduced in this context in~\cite{Hannestad:2000gt} (and first used for a treatment of self-interactions in light neutrinos in~\cite{Hannestad:2004qu}) in which the scattering interactions lead to damping of all terms in the Boltzmann hierarchy beyond $\ell = 1$. RTA was found to be very accurate and consistent with the exact description of Boltzmann equations, in~\cite{Oldengott:2017fhy}. The collisional Boltzmann hierarchy for massive neutrinos used in \CAMB{} in synchronous gauge (following the notation in \cite{Ma:1995ey}) can then be written as, 
\begin{eqnarray}
\label{eq:boltzman}
\dot{\Psi}_0 &=& -{qk\over \epsilon}\Psi_1
+{1\over 6}\dot{h} {d\ln f_0\over d\ln q}
\,, \nonumber\\
\dot{\Psi}_1 &=& {qk\over 3\epsilon} \left(\Psi_0
- 2 \Psi_2 \right) \,, \nonumber\\
\dot{\Psi}_2 &=& {qk\over 5\epsilon} \left(
2\Psi_1 - 3\Psi_3 \right)
- \left( {1\over15}\dot{h} + {2\over5} \dot{\eta} \right)
{d\ln f_0\over d\ln q} + \alpha_2 \dot{\tau}_\nu \Psi_2\,,\\
\dot{\Psi}_l &=& {qk \over (2l+1)\epsilon} \left[ l\Psi_{l-1}
- (l+1)\Psi_{l+1} \right] + \alpha_\ell \dot{\tau}_\nu \Psi_l \,,
\quad l \geq 3 \,. \nonumber
\end{eqnarray}

where $\dot \tau_\nu \equiv -a G_{\rm eff}^2 T_\nu^5$ is the neutrino self-interaction opacity, and $\alpha_l$ ($l>1$) are model dependent coefficients of order unity. We use $\alpha_l$ values given by equation\ 2.9 in~\cite{Oldengott:2017fhy} for the flavour-independent diagonal interaction. Specifically, we use $\alpha_2 = 0.40$, $\alpha_3 = 0.43$, $\alpha_4 = 0.46$,  $\alpha_5 = 0.47$, $\alpha_{l \geq 6} = 0.48$ (here we again note that the specific numbers were derived for a scalar mediator in \cite{Oldengott:2017fhy}, but that assuming pseudoscalar or vector interactions will lead to almost identical results). 
Since the equations~\ref{eq:boltzman} are difficult to solve in the very early universe, we use a tight coupling approximation (TCA) where only the two lowest moments are non-zero. We switch the TCA off when $|\dot \tau_\nu|/\mathcal{H} < 100$, where $\mathcal{H}$ is the conformal Hubble parameter. We have verified that this does not bias our results.

\subsection{Cosmological model: parameterisation and priors}

We perform our analysis in the framework of the $\Lambda$CDM model, including massive neutrinos with a mass sum $\sum m_{\nu}$, and interactions between them parameterised in terms of $G_{\rm eff}$, as well as the effective number of relativistic degrees of freedom, $N_{\rm eff}$.

The self-interacting neutrino cosmological model is therefore fully specified by the parameter vector
\begin{equation}\label{eq:model}
{\bm \theta} = \{\Omega_{\rm c} h^2,\Omega_{\rm b} h^2,100\theta_{MC},\tau,
\ln(10^{10}A_{s}),n_{s}, \sum m_\nu, N_{\rm eff}, \textrm{log}_{10} \left[G_{\rm eff} \textrm{MeV}^2\right]\}.
\end{equation}
Here, $\Omega_{\rm c} h^2$ and $\Omega_{\rm b} h^2$ are the present-day physical CDM and baryon densities respectively, $\theta_{MC}$ the parameter used by CosmoMC to parameterise the angular size of the sound horizon, $\tau$  the optical depth to reionization, and $\ln(10^{10}A_s)$ and $n_s$ denote respectively the amplitude and spectral index of the initial scalar fluctuations.
We assume that neutrino masses are degenerate so that $m_\nu = \frac{1}{3}\sum m_\nu$.
For the interaction strength, $G_{\rm eff}$, we use a prior which is flat in $\textrm{log}_{10} \left[G_{\rm eff} \textrm{MeV}^2\right]$, where $G_{\rm eff}$ is expressed in units of MeV$^{-2}$. While this is not optimal in the sense of being a Jeffreys prior, it allows us to vary the unknown $G_{\rm eff} $ over multiple orders of magnitude.

The uniform priors applied on the cosmological parameters are given in table~\ref{tab:priors}. Since the posterior of $\textrm{log}_{10} \left[G_{\rm eff} \textrm{MeV}^2\right]$ for the given prior range in table~\ref{tab:priors} is bimodal, it is difficult to obtain Bayesian evidence and parameter constraints for the MI$\nu$ and the SI$\nu$ modes separately from the runs incorporating the full prior range of $\textrm{log}_{10} \left[G_{\rm eff} \textrm{MeV}^2\right]$. To do that, we run separate nested-sampling analysis for each mode by splitting the prior range of $\textrm{log}_{10} \left[G_{\rm eff} \textrm{MeV}^2\right]$. For all runs including the Planck 2018 CMB high-$l$ polarisation data, we use a prior range of $-5.5\rightarrow -2.3$ for the MI$\nu$, and $-2.3\rightarrow -0.1$ for the SI$\nu$, all other parameter prior ranges remaining unchanged. For runs without the Planck 2018 high-$l$ polarisation data, the prior ranges are $-5.5\rightarrow -2$ and $-2\rightarrow -0.1$ for the MI$\nu$ and SI$\nu$ respectively.

To compare with the case of non-interacting massive neutrinos, we also perform analysis in the $\Lambda\rm CDM + N_{\rm eff}+ \sum m_{\nu}$ model. We denote this case by NI$\nu$.

\begin{table}[t]
\caption{Priors for the cosmological parameters considered. All priors are uniform.}
\label{tab:priors}
\begin{center}
\begin{tabular}{lr@{$\,\to\,$}l}
\hline
 Parameter & \multicolumn{2}{c}{Prior}\\
\hline
$\Omega_{\rm b}h^2$ & $0.019$ & $0.025$\\
$\Omega_{\rm c}h^2$ & $0.095$ & $0.145$\\
$100\theta_{MC}$ & $1.03$ & $1.05$\\
$\tau$ & $0.01$ & $0.1$\\
$n_s$ & $0.885$ & $1.04$\\
$\ln{(10^{10} A_s)}$ & $2.5$ & $3.7$\\
$\sum m_\nu$ [eV] &  $0.005$ & $1$\\
$N_{\rm eff}$ & $2$ & $5$\\
$\textrm{log}_{10} \left[G_{\rm eff} \textrm{MeV}^2\right]$  & $-5.5$ & $-0.1$\\
\hline
\end{tabular}
\end{center}
\end{table}

\subsection{Datasets}\label{sec:datasets}

The main data used is the CMB data from the Planck 2018 data release~\cite{Aghanim:2018eyx}. We use these both without and with high-$\ell$ polarisation (TT and TTTEEE respectively), but always including low-$\ell$ E mode polarisation (lowE). Specifically, TT denotes the low-$l$ and high-$l$ temperature power spectra, whereas TTTEEE denotes the TT spectra combined with high-$l$ TE and EE spectra. Also note that in table \ref{tab:evidence}, for convenience, we also use the abbreviation ``CMB'' for TTTEEE+lowE.

In addition to the primary data set, we use an auxiliary dataset combination denoted by EXT which consists of Planck 2018 CMB lensing~\cite{Aghanim:2018oex}, BAO and RSD measurements from SDSS-III BOSS DR12~\cite{Alam:2016hwk}, additional BAO measurements from MGS \cite{Ross:2014qpa} and 6dFGS \cite{Beutler:2011hx}, and SNe~Ia luminosity distance measurements from the Pantheon sample~\cite{Scolnic:2017caz}. 
Finally, in order to be able to comment on the effect of neutrino non-standard interactions on the current Hubble discrepancy we also combine the CMB data with the local $H_0$ measurement from ref.~\cite{Riess:2019cxk} which we denote R19.

\subsection{Parameter sampling and analysis}

We sample the parameter space using the PolyChord~\cite{Handley:2015vkr,Handley:2015fda} extension of CosmoMC~\cite{Lewis:2002ah,Lewis:2013hha}, called CosmoChord \cite{Handley}, where we modify the \CAMB{} code to incorporate the neutrino self-interactions. Use of the nested-sampling package PolyChord enables us to calculate evidences accurately, and properly sample this parameter space of bimodal posterior distributions. For the runs which incorporated the full prior range of $\textrm{log}_{10} \left[G_{\rm eff} \textrm{MeV}^2\right]$, i.e. $-5.5$ $\rightarrow$ $-0.1$, we used 4000 live points with the setting boost\_posterior = 3, to ensure that our analyses lead to accurate posterior distribution of the cosmological parameters. As stated before, we also run the non-interacting case (NI$\nu$), the moderately interacting case (MI$\nu$), and the strongly interacting case (SI$\nu$) separately. Since the parameter posterior distributions for NI$\nu$, MI$\nu$, and SI$\nu$ are unimodal, for these analyses we use a less computationally demanding setting of 2000 live points and boost\_posterior = 0. We use HMcode \cite{Mead:2020vgs} (included with the CosmoChord package) to handle non-linearities. The posterior distributions and parameter constraints were derived using GetDist~\cite{Lewis:2019xzd}.

\section{Results}
\label{section:3}

\begin{figure}[tbp]
    \includegraphics[width=\textwidth]{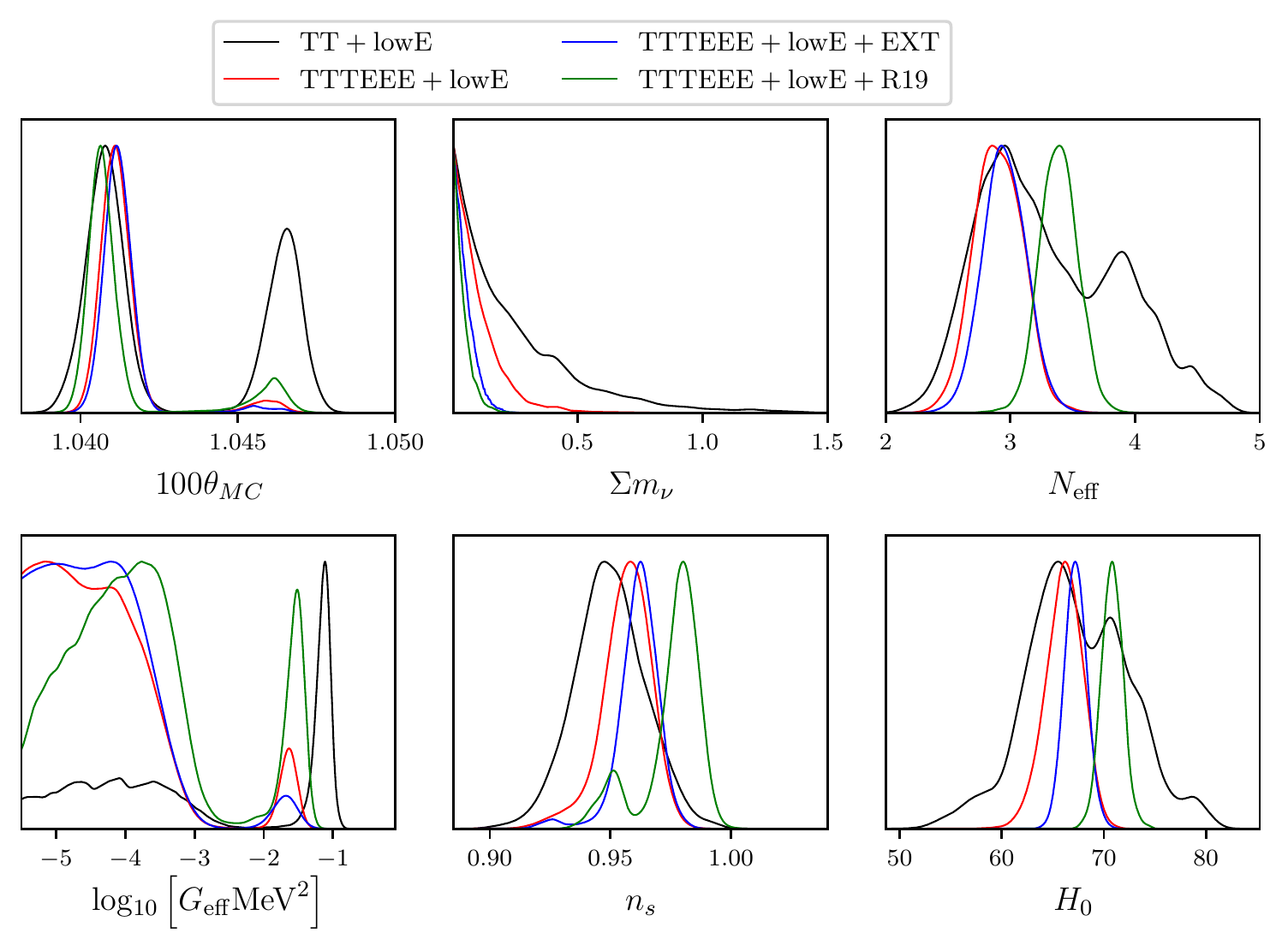}
    \caption{\label{fig:posteriorsFullRange} Posterior distributions for the full range of coupling strengths $\logg$. Using just temperature data, the posterior for $\logg$ is clearly multimodal with a second peak at $\logg\simeq -1.2$. When adding more data, the interacting mode is shifted slightly to the left and the peak generally becomes less pronounced. The exception is the addition of the R19 data which enhances the second peak.}
\end{figure}

\begin{figure}[tbp]
    \includegraphics[width=\textwidth]{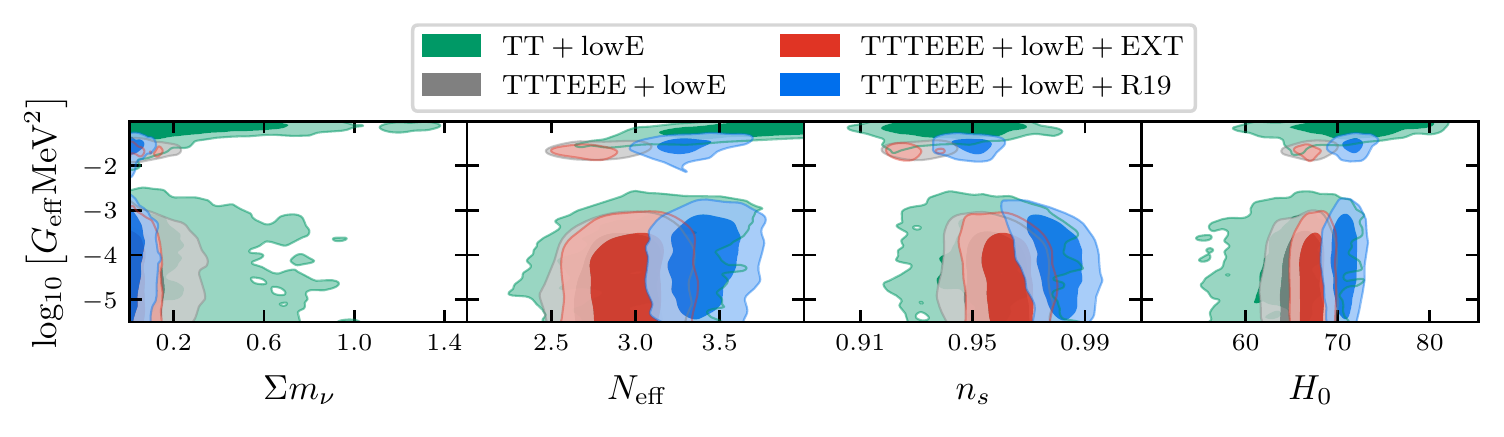}
    \caption{\label{fig:contoursFullRange} Two-dimensional posterior distributions for the full range of coupling strengths $\logg$. Considering just temperature data and large couplings, $\logg \gtrsim -1.5$, we not that there is a large allowed volume in parameter space stretching towards large values of $\mnu$ and $\nnu$. Adding more data constrains these parameters which effectively shifts the peak towards smaller values of $\logg$ as seen in figure~\ref{fig:posteriorsFullRange}.}
\end{figure}

\begin{figure}[tbp]
\includegraphics[width=\textwidth]{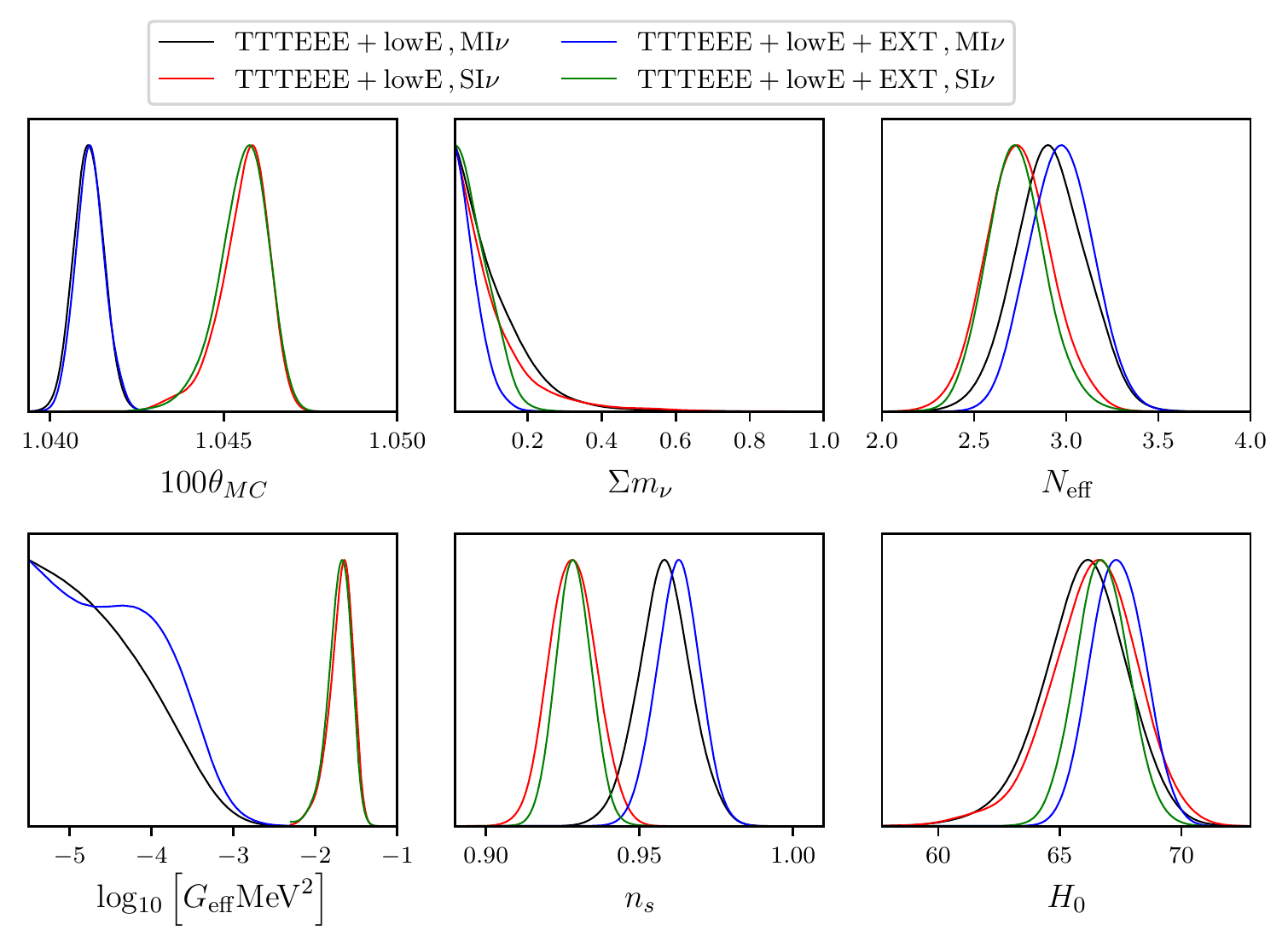}
\caption{\label{fig:posteriorsCMBCMBext} Posterior distributions for the moderately interacting mode ($\MIn$) and the strongly interacting mode ($\SIn$). The $\SIn$-mode is associated with a notably smaller value for scalar spectral index $n_s$ and a significantly larger value of $\theta_\text{MC}$. The central value of $\nnu$ is also slightly smaller in the $\SIn$-mode. Adding the extra datasets (EXT) tightens the mass-bound and removes the low-$H_0$ tail, and this effect is common to both modes.}
\end{figure}

\begin{figure}[tbp]
    \includegraphics[width=\textwidth]{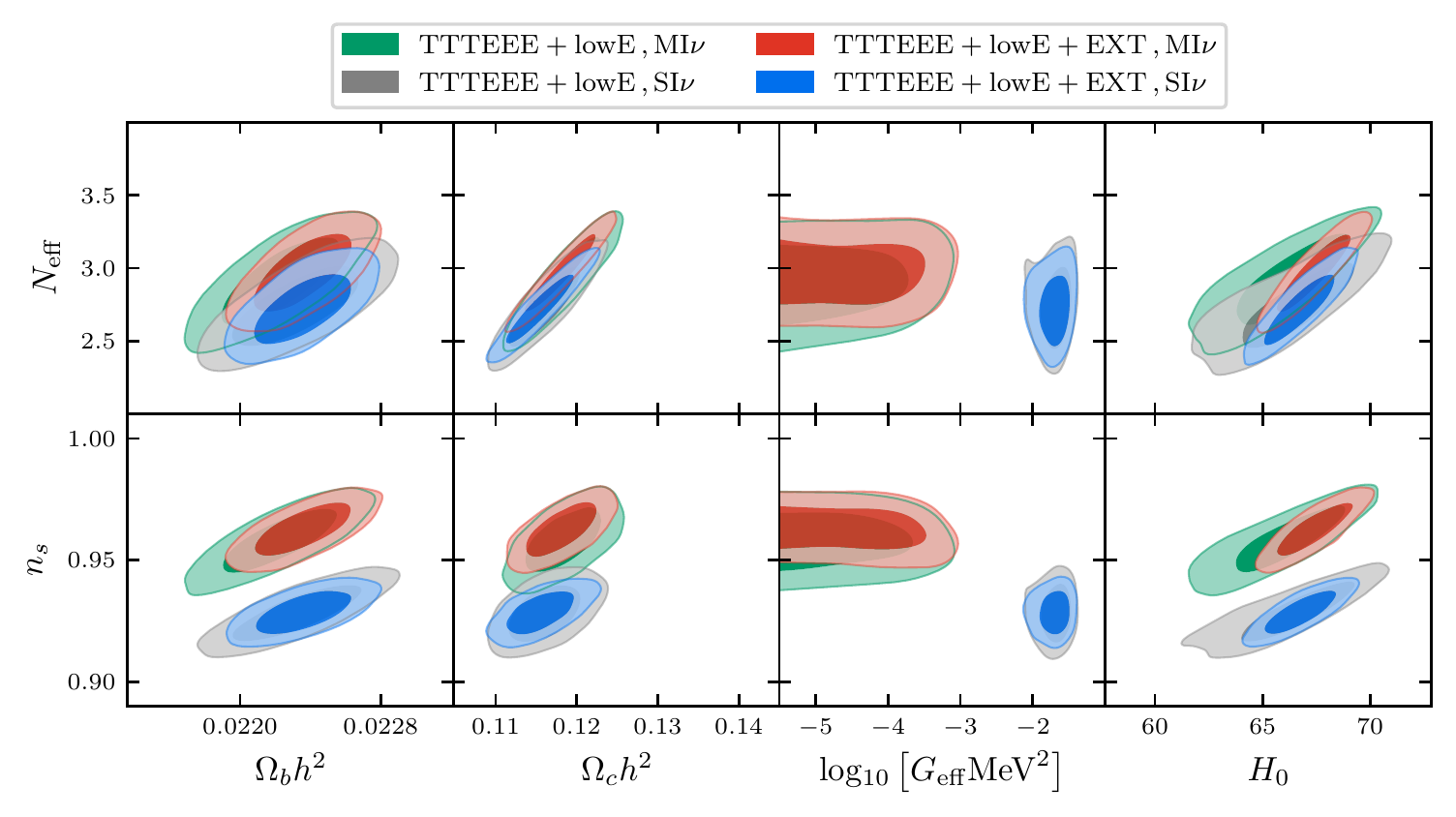}
    \caption{\label{fig:contoursCMBCMBext} Two-dimensional posterior distributions for the moderately interacting mode ($\MIn$) and the strongly interacting mode ($\SIn$). Adding the external data primarily affects the allowed $H_0-\nnu$-posterior.
    Considering just temperature data and large couplings, $\logg \gtrsim -1.3$, we not that there is a large allowed volume in parameter space stretching towards large values of $\mnu$ and $\nnu$. Adding more data constrains these parameters which effectively shifts the peak towards smaller values of $\logg$ as seen in figure~\ref{fig:posteriorsFullRange}.}
\end{figure}

\begin{table}
\renewcommand*{\arraystretch}{1.2}
\begin{tabular} { l r|  c c c c}

\hline
& & TT+lowE &  CMB &  CMB+EXT  &  CMB+R19\\
\hline
\hline
\multirow{3}{*}{{\boldmath$\Omega_b h^2   $} }&NI$\nu$& $0.02194^{+0.00072}_{-0.00078}$ & $0.02223^{+0.00044}_{-0.00043}$ & $0.02235^{+0.00037}_{-0.00037}$ & $0.02276^{+0.00035}_{-0.00035}$\\
&MI$\nu$& $0.02191^{+0.00070}_{-0.00078}$ & $0.02222^{+0.00043}_{-0.00043}$ & $0.02236^{+0.00035}_{-0.00035}$ & $0.02278^{+0.00035}_{-0.00033}$\\
&SI$\nu$& $0.02226^{+0.00074}_{-0.00081}$ & $0.02232^{+0.00046}_{-0.00045}$ & $0.02236^{+0.00035}_{-0.00035}$ & $0.02287^{+0.00033}_{-0.00034}$\\

\hline
\multirow{3}{*}{{\boldmath$\Omega_c h^2   $} }&NI$\nu$& $0.1200^{+0.0078}_{-0.0079}$ & $0.1183^{+0.0060}_{-0.0057}$ & $0.1179^{+0.0055}_{-0.0056}$ & $0.1234^{+0.0057}_{-0.0056}$\\
&MI$\nu$& $0.1201^{+0.0086}_{-0.0080}$ & $0.1184^{+0.0061}_{-0.0059}$ & $0.1182^{+0.0055}_{-0.0055}$ & $0.1238^{+0.0057}_{-0.0055}$\\
&SI$\nu$& $> 0.126                   $ & $0.1161^{+0.0061}_{-0.0057}$ & $0.1156^{+0.0060}_{-0.0052}$ & $0.1220^{+0.0065}_{-0.0060}$\\

\hline
\multirow{3}{*}{{\boldmath$\Sigma m_\nu   $} }&NI$\nu$& $< 0.705                   $ & $< 0.297                   $ & $< 0.122                   $ & $< 0.105                   $\\
&MI$\nu$& $< 0.771                   $ & $< 0.290                   $ & $< 0.117                   $ & $< 0.0917                  $\\
&SI$\nu$& $< 0.848                   $ & $< 0.325                   $ & $< 0.152                   $ & $< 0.105                   $\\

\hline
\multirow{3}{*}{{\boldmath$N_\mathrm{eff} $} }&NI$\nu$& $2.95^{+0.59}_{-0.59}      $ & $2.91^{+0.39}_{-0.37}      $ & $2.96^{+0.33}_{-0.35}      $ & $3.38^{+0.30}_{-0.30}      $\\
&MI$\nu$& $2.96^{+0.61}_{-0.59}      $ & $2.91^{+0.38}_{-0.38}      $ & $2.97^{+0.34}_{-0.33}      $ & $3.41^{+0.31}_{-0.30}      $\\
&SI$\nu$& $4.00^{+0.80}_{-0.82}      $ & $2.74^{+0.38}_{-0.35}      $ & $2.73^{+0.34}_{-0.31}      $ & $3.22^{+0.32}_{-0.30}      $\\

\hline
\multirow{3}{*}{{\boldmath$\log_{10} G_\mathrm{eff}$} }&NI$\nu$&                              &                              &                              &                             \\
&MI$\nu$& $< -3.04                   $ & $< -3.47                   $ & $< -3.37                   $ & $< -3.27                   $\\
&SI$\nu$& $-1.13^{+0.20}_{-0.21}     $ & $-1.69^{+0.27}_{-0.31}     $ & $-1.71^{+0.27}_{-0.31}     $ & $-1.58^{+0.29}_{-0.37}     $\\

\hline
\multirow{3}{*}{{\boldmath$n_s            $} }&NI$\nu$& $0.957^{+0.029}_{-0.030}   $ & $0.959^{+0.017}_{-0.017}   $ & $0.963^{+0.014}_{-0.014}   $ & $0.981^{+0.012}_{-0.013}   $\\
&MI$\nu$& $0.954^{+0.028}_{-0.031}   $ & $0.958^{+0.017}_{-0.017}   $ & $0.963^{+0.014}_{-0.014}   $ & $0.981^{+0.013}_{-0.013}   $\\
&SI$\nu$& $0.944^{+0.028}_{-0.029}   $ & $0.928^{+0.015}_{-0.015}   $ & $0.929^{+0.012}_{-0.011}   $ & $0.947^{+0.011}_{-0.011}   $\\

\hline
\multirow{3}{*}{$H_0                       $ }&NI$\nu$& $64.6^{+6.3}_{-8.2}        $ & $65.9^{+3.3}_{-3.8}        $ & $67.3^{+2.2}_{-2.2}        $ & $70.5^{+2.1}_{-2.0}        $\\
&MI$\nu$& $64.6^{+7.0}_{-8.1}        $ & $66.0^{+3.5}_{-3.6}        $ & $67.4^{+2.2}_{-2.1}        $ & $70.7^{+2.2}_{-2.1}        $\\
&SI$\nu$& $73^{+9}_{-10}             $ & $66.4^{+3.7}_{-3.7}        $ & $66.7^{+2.2}_{-2.1}        $ & $71.0^{+2.2}_{-2.1}        $\\
\hline
\hline
\multirow{3}{*}{$-2 \left[\log \left( \mathcal{L} / \mathcal{L}_{\text{NI}\nu} \right) \right]$ \kern-1em} & NI$\nu$ & - & - & - & - \\
&MI$\nu$& -1.0& -1.2& 0.2& -1.1\\
&SI$\nu$& -2.9& 3.0& 3.4& 0.8\\
\hline
\multirow{3}{*}{$Z/Z_{\text{NI}\nu}$} & NI$\nu$ & - & - & - & - \\
&MI$\nu$& 0.67& 0.47& 0.45& 0.47\\
&SI$\nu$& 1.30& 0.03& 0.06& 0.06\\
\hline\hline
\end{tabular}

\caption{\label{tab:evidence} Parameter constraints (95\%) in the non-interacting model (NI$\nu$), moderately interacting model (MI$\nu$) and the strongly interacting model (SI$\nu$). The constraints are reported for four different dataset combinations: TT+lowE, TTTEEE+lowE (abbreviated CMB), CMB+EXT and CMB+R19. EXT refers to the extra dataset combinations defined in section~\ref{sec:datasets} and R19 refers to the local measurement of the Hubble constant. For each dataset combination we have reported the ratio of Bayesian evidences and the difference in bestfit log-likelihoods w.r.t. the non-interacting case NI$\nu$.}
\end{table}

Our main results from the cosmological parameter estimation runs are tabulated in table~\ref{tab:evidence}, and visualised in figure~\ref{fig:posteriorsFullRange}--\ref{fig:contoursCMBCMBext}.

We can briefly summarise our results regarding the cosmological parameters in the following way:

\begin{itemize}

\item
$G_{\rm eff}$: When we use the full parameter range and PolyChord we find the two-peak structure previously established in literature. However, due to lack of sampling there is no relevant constraint on $G_{\rm eff}$. Once we split parameter space into the MI$\nu$ and SI$\nu$ ranges we find that the SI$\nu$ allowed parameter range is centered around $\log G_{\rm eff} \sim -1.7 \to -1.1$, roughly consistent with previous findings. We also find that the allowed range does not depend strongly on the data sets used.

\item
$\sum m_\nu$: We find that although the SI$\nu$ range prefers slightly larger values of $m_\nu$ than the MI$\nu$ range, the inferred upper limit on the neutrino mass does not change significantly (typically less than 20--30\%). This means that the cosmological neutrino mass bounds quoted in literature are quite robust against the type of changes to the neutrino sector studied in this work.

\item
$N_{\rm eff}$: Here we find that for the SI$\nu$ range the preferred value is systematically lower than for the MI$\nu$/NI$\nu$ case. However, the preferred range does not shift by more than $1\sigma$, and $N_{\rm eff}=3.046$ is in all cases well within the allowed range. The only exception to this is when the R19 $H_0$ data is used. In this particular case, both the MI and SI preferred ranges for $N_{\rm eff}$ shift to values larger than 3 because of the well known correlation between $H_0$ and $N_{\rm eff}$ also seen in the upper right corner of figure 4.

\item
$n_s$: Here we also recover a result previously found in literature, namely that the SI$\nu$ region shifts the preferred range of $n_s$ to significantly lower values, with potential implications for the validity of a number of inflation models (see~\cite{Barenboim:2019tux} for a more detailed discussion of this point).

\item
$H_0$: The possibility of neutrino self-interactions has previously been invoked as a possible solution to the $H_0$ discrepancy. For example,~\cite{Kreisch:2019yzn} found that significantly higher values of $H_0$ could be allowed in their analysis of Planck 2015 data when neutrino self-interactions were included (see also e.g.~\cite{Archidiacono:2020yey} for a discussion of this issue in the context of models with sterile neutrinos). However, we find no evidence that significantly higher $H_0$ values can be allowed by adding neutrino self-interactions. In fact we find that all limits on $H_0$ are close to independent of $G_{\rm eff}$. The only case where we find preference for a higher value is when the R19 data is used, and in that case it is purely driven by the data added, not by the model.
The difference in our finding compared to~\cite{Kreisch:2019yzn} most likely is that they have not used low-$\ell$ polarisation data. This leads to a much looser bound on neutrino parameters and subsequently also on $H_0$.

\end{itemize}

We should note that our choice of using a logarithmic prior in $G_\text{eff}$ directly impacts the relative normalisation of the two peaks in figure~\ref{fig:posteriorsFullRange}. From the transformation theorem we see that switching to a linear prior in $G_\text{eff}$ amounts to multiplying the posterior in figure~\ref{fig:posteriorsFullRange} by $G_\text{eff}$. Since the separation of the peaks in $G_\text{eff}$ is roughly $10^2$, the height of the second peak would have been much higher than the first peak. However, this would be quite misleading since the quality of the fit at the second peak is in general worse than the first peak as we shall discuss next.

\subsection{Bayesian evidence and model selection}

Next, we discuss the relative evidence for and against the MI$\nu$ and SI$\nu$ peaks, relative to the standard $\Lambda$CDM model. We use two measures:
\begin{itemize}
    \item The ratio $Z/Z_{\text{NI}\nu}$, i.e. the Bayesian evidence for the interacting models (MI$\nu$ and SI$\nu$) divided by the Bayesian evidence of the non-interacting models.
    \item The difference in log-likelihoods between the interacting models and the non-interacting models at their respective bestfit points.
\end{itemize}
The Bayesian evidence is usually the preferred statistic for model selection since it penalises highly tuned models. However, for the same reason it is also very sensitive to the choice of prior, and our choice of using a logarithmic prior in $G_{\rm eff}$ penalises large couplings to some extent. In this case then, where the evidence is highly sensitive to the prior, the log-likelihood at the bestfit point can be used as an additional argument for or against the model.

Looking at table~\ref{tab:evidence} we see that only one dataset combination (TT+lowE) leads to an evidence ratio larger than 1 for the SI$\nu$-model, and only slightly so. For the rest of the dataset combinations SI$\nu$ is strongly disfavoured (according to Jeffrey's scale~\cite{jeffreys1998theory}) compared to the non-interacting model. MI$\nu$ is in all cases mildly disfavoured compared to the non-interacting model.

We also note that since the NI$\nu$ parameter space is a proper subset of the MI$\nu$ parameter space, we should expect $-2 \log \mathcal{L}_{\text{MI}\nu} < -2 \log \mathcal{L}_{\text{NI}\nu}$ up to numerical uncertainties, and that is indeed what we find. To judge if the improvement is sufficient to warrant the introduction of another parameter, we may use the Akaike information criterion (AIC)~\cite{Akaike1975},
\begin{equation}
    \text{AIC} = 2k - 2\log\mathcal{L}\,,
\end{equation}
where $k$ is the number of parameters in the model. We find
\begin{align}
    \text{AIC}_{\text{MI}\nu} - \text{AIC}_{\text{NI}\nu} &= 2 + \left(- 2\left[ \log\mathcal{L}_{\text{MI}\nu} - \log\mathcal{L}_{\text{NI}\nu} \right] \right)\,
\end{align}
where the parenthesis is the quantity shown in table~\ref{tab:evidence}. Comparing with the numbers in table~\ref{tab:evidence}, we find that the non-interacting model is preferred to the interacting model except in the dataset combination TT+lowE. The values can be interpreted such that the interacting model is roughly 65\% as probable as the non-interacting model to minimise information loss for the dataset combinations TTTEEE+lowE and TTTEEE+lowE+R19 and about 35\% for the CMB+EXT data.

\section{Conclusions}
\label{section:4}
We have studied constraints on possible non-standard neutrino interactions with a heavy scalar mediator, in the effective 4-fermion interaction limit, from cosmological observables, using the well-established relaxation time approximation to modify the collisional neutrino Boltzmann equations. Specifically we have used the latest CMB temperature, polarisation, and lensing data from the Planck 2018 data release, BAO and RSD measurements from SDSS-III BOSS DR-12, additional BAO measurements from MGS and 6dFGS, and uncalibrated SNe~Ia luminosity distance data from the Pantheon sample.

Consistent with earlier findings we find that albeit observables prefer the standard $\Lambda$CDM model over models with significant self-interactions, there is a region of large $G_\text{eff}$ which continues to provide a reasonable fit to data. This limits the ability of structure formation to stringently constrain $G_\text{eff}$, and could in principle point to the presence of non-standard physics in the neutrino sector. 
Reassuringly, we find that the neutrino self-interactions studied here leave the cosmological neutrino mass bound almost unchanged so that the mass bounds found in the literature are robust against this particular type of new physics.  Finally, consistent with earlier results, we find that strong neutrino self-interactions leads to lower values of $N_{\rm eff}$ being preferred, although not at high significance, and that the preferred value of $n_s$ shifts to significantly lower values. 

However, we also find that with current data, the strongly interacting mode is not favoured. In fact, we find that once high-$\ell$ polarisation from Planck 2018 is used the strongly interacting mode is always disfavoured, both in terms of Bayesian evidence, {\it and} from the raw likelihood.
We also find that the strongly interacting mode no longer provides a viable solution to the $H_0$ tension, except when high-$\ell$ polarisation is ignored, making it substantially less interesting. 
Finally, we note that the although the results presented here were derived for the case of a heavy scalar mediator, they also apply to e.g.\ the case of a vector mediator with minor shifts in preferred values of $G_{\rm eff}$.

\section*{Acknowledgements}
We acknowledge the use of the HPC facility at Aarhus University (http://www.cscaa.dk/) for all the numerical analyses done in this work. T.T.  was supported by
a research grant (29337) from VILLUM FONDEN.

\appendix

\bibliographystyle{utcaps}
\bibliography{references}

\providecommand{\href}[2]{#2}\begingroup\raggedright\begin{thebibliography}{10}

\bibitem{Fukuda:1998mi}
{\bfseries Super-Kamiokande} Collaboration, Y.~Fukuda {\em et~al.}, ``{Evidence
  for oscillation of atmospheric neutrinos},''
  \href{http://dx.doi.org/10.1103/PhysRevLett.81.1562}{{\em Phys. Rev. Lett.}
  {\bfseries 81} (1998) 1562--1567},
\href{http://arxiv.org/abs/hep-ex/9807003}{{\ttfamily arXiv:hep-ex/9807003
  [hep-ex]}}.

\bibitem{Ahmad:2002jz}
{\bfseries SNO} Collaboration, Q.~R. Ahmad {\em et~al.}, ``{Direct evidence for
  neutrino flavor transformation from neutral current interactions in the
  Sudbury Neutrino Observatory},''
  \href{http://dx.doi.org/10.1103/PhysRevLett.89.011301}{{\em Phys. Rev. Lett.}
  {\bfseries 89} (2002) 011301},
\href{http://arxiv.org/abs/nucl-ex/0204008}{{\ttfamily arXiv:nucl-ex/0204008
  [nucl-ex]}}.

\bibitem{Aghanim:2018eyx}
{\bfseries Planck} Collaboration, N.~Aghanim {\em et~al.}, ``{Planck 2018
  results. VI. Cosmological parameters},''
  \href{http://dx.doi.org/10.1051/0004-6361/201833910}{{\em Astron. Astrophys.}
  {\bfseries 641} (2020) A6}, \href{http://arxiv.org/abs/1807.06209}{{\ttfamily
  arXiv:1807.06209 [astro-ph.CO]}}.

\bibitem{RoyChoudhury:2019hls}
S.~Roy~Choudhury and S.~Hannestad, ``{Updated results on neutrino mass and mass
  hierarchy from cosmology with Planck 2018 likelihoods},''
  \href{http://dx.doi.org/10.1088/1475-7516/2020/07/037}{{\em JCAP} {\bfseries
  2007} (2020) 037},
\href{http://arxiv.org/abs/1907.12598}{{\ttfamily arXiv:1907.12598
  [astro-ph.CO]}}.

\bibitem{Choudhury:2018byy}
S.~Roy~Choudhury and S.~Choubey, ``{Updated Bounds on Sum of Neutrino Masses in
  Various Cosmological Scenarios},''
  \href{http://dx.doi.org/10.1088/1475-7516/2018/09/017}{{\em JCAP} {\bfseries
  1809} (2018) 017},
\href{http://arxiv.org/abs/1806.10832}{{\ttfamily arXiv:1806.10832
  [astro-ph.CO]}}.

\bibitem{Alam:2020sor}
{\bfseries eBOSS} Collaboration, S.~Alam {\em et~al.}, ``{The Completed SDSS-IV
  extended Baryon Oscillation Spectroscopic Survey: Cosmological Implications
  from two Decades of Spectroscopic Surveys at the Apache Point observatory},''
\href{http://arxiv.org/abs/2007.08991}{{\ttfamily arXiv:2007.08991
  [astro-ph.CO]}}.

\bibitem{Vagnozzi:2017ovm}
S.~Vagnozzi, E.~Giusarma, O.~Mena, K.~Freese, M.~Gerbino, S.~Ho, and
  M.~Lattanzi, ``{Unveiling $\nu$ secrets with cosmological data: neutrino
  masses and mass hierarchy},''
  \href{http://dx.doi.org/10.1103/PhysRevD.96.123503}{{\em Phys. Rev. D}
  {\bfseries 96} no.~12, (2017) 123503},
  \href{http://arxiv.org/abs/1701.08172}{{\ttfamily arXiv:1701.08172
  [astro-ph.CO]}}.

\bibitem{Gelmini:1980re}
G.~B. Gelmini and M.~Roncadelli, ``{Left-Handed Neutrino Mass Scale and
  Spontaneously Broken Lepton Number},''
\href{http://dx.doi.org/10.1016/0370-2693(81)90559-1}{{\em Phys. Lett.}
  {\bfseries 99B} (1981) 411--415}.

\bibitem{Choi:1991aa}
K.~Choi and A.~Santamaria, ``{17-KeV neutrino in a singlet - triplet majoron
  model},''
\href{http://dx.doi.org/10.1016/0370-2693(91)90900-B}{{\em Phys. Lett.}
  {\bfseries B267} (1991) 504--508}.

\bibitem{Acker:1992eh}
A.~Acker, A.~Joshipura, and S.~Pakvasa, ``{A Neutrino decay model, solar
  anti-neutrinos and atmospheric neutrinos},''
\href{http://dx.doi.org/10.1016/0370-2693(92)91520-J}{{\em Phys. Lett.}
  {\bfseries B285} (1992) 371--375}.

\bibitem{Chikashige:1980ui}
Y.~Chikashige, R.~N. Mohapatra, and R.~D. Peccei, ``{Are There Real Goldstone
  Bosons Associated with Broken Lepton Number?},''
\href{http://dx.doi.org/10.1016/0370-2693(81)90011-3}{{\em Phys. Lett.}
  {\bfseries 98B} (1981) 265--268}.

\bibitem{Georgi:1981pg}
H.~M. Georgi, S.~L. Glashow, and S.~Nussinov, ``{Unconventional Model of
  Neutrino Masses},''
\href{http://dx.doi.org/10.1016/0550-3213(81)90336-9}{{\em Nucl. Phys.}
  {\bfseries B193} (1981) 297--316}.

\bibitem{Oldengott:2017fhy}
I.~M. Oldengott, T.~Tram, C.~Rampf, and Y.~Y. Wong, ``{Interacting neutrinos in
  cosmology: exact description and constraints},''
  \href{http://dx.doi.org/10.1088/1475-7516/2017/11/027}{{\em JCAP} {\bfseries
  11} (2017) 027}, \href{http://arxiv.org/abs/1706.02123}{{\ttfamily
  arXiv:1706.02123 [astro-ph.CO]}}.

\bibitem{Oldengott:2014qra}
I.~M. Oldengott, C.~Rampf, and Y.~Y.~Y. Wong, ``{Boltzmann hierarchy for
  interacting neutrinos I: formalism},''
  \href{http://dx.doi.org/10.1088/1475-7516/2015/04/016}{{\em JCAP} {\bfseries
  1504} (2015) 016},
\href{http://arxiv.org/abs/1409.1577}{{\ttfamily arXiv:1409.1577
  [astro-ph.CO]}}.

\bibitem{Barenboim:2019tux}
G.~Barenboim, P.~B. Denton, and I.~M. Oldengott, ``{Constraints on inflation
  with an extended neutrino sector},''
  \href{http://dx.doi.org/10.1103/PhysRevD.99.083515}{{\em Phys. Rev.}
  {\bfseries D99} no.~8, (2019) 083515},
\href{http://arxiv.org/abs/1903.02036}{{\ttfamily arXiv:1903.02036
  [astro-ph.CO]}}.

\bibitem{Kreisch:2019yzn}
C.~D. Kreisch, F.-Y. Cyr-Racine, and O.~Dor\'e, ``{Neutrino puzzle: Anomalies,
  interactions, and cosmological tensions},''
  \href{http://dx.doi.org/10.1103/PhysRevD.101.123505}{{\em Phys. Rev. D}
  {\bfseries 101} no.~12, (2020) 123505},
  \href{http://arxiv.org/abs/1902.00534}{{\ttfamily arXiv:1902.00534
  [astro-ph.CO]}}.

\bibitem{Archidiacono:2013dua}
M.~Archidiacono and S.~Hannestad, ``{Updated constraints on non-standard
  neutrino interactions from Planck},''
  \href{http://dx.doi.org/10.1088/1475-7516/2014/07/046}{{\em JCAP} {\bfseries
  1407} (2014) 046},
\href{http://arxiv.org/abs/1311.3873}{{\ttfamily arXiv:1311.3873
  [astro-ph.CO]}}.

\bibitem{Forastieri:2019cuf}
F.~Forastieri, M.~Lattanzi, and P.~Natoli, ``{Cosmological constraints on
  neutrino self-interactions with a light mediator},''
  \href{http://dx.doi.org/10.1103/PhysRevD.100.103526}{{\em Phys. Rev. D}
  {\bfseries 100} no.~10, (2019) 103526},
  \href{http://arxiv.org/abs/1904.07810}{{\ttfamily arXiv:1904.07810
  [astro-ph.CO]}}.

\bibitem{Archidiacono:2020yey}
M.~Archidiacono, S.~Gariazzo, C.~Giunti, S.~Hannestad, and T.~Tram, ``{Sterile
  neutrino self-interactions: $H_0$ tension and short-baseline anomalies},''
  \href{http://dx.doi.org/10.1088/1475-7516/2020/12/029}{{\em JCAP} {\bfseries
  12} (2020) 029}, \href{http://arxiv.org/abs/2006.12885}{{\ttfamily
  arXiv:2006.12885 [astro-ph.CO]}}.

\bibitem{Blinov:2019gcj}
N.~Blinov, K.~J. Kelly, G.~Z. Krnjaic, and S.~D. McDermott, ``{Constraining the
  Self-Interacting Neutrino Interpretation of the Hubble Tension},''
  \href{http://dx.doi.org/10.1103/PhysRevLett.123.191102}{{\em Phys. Rev.
  Lett.} {\bfseries 123} no.~19, (2019) 191102},
\href{http://arxiv.org/abs/1905.02727}{{\ttfamily arXiv:1905.02727
  [astro-ph.CO]}}.

\bibitem{Ohlsson:2012kf}
T.~Ohlsson, ``{Status of non-standard neutrino interactions},''
  \href{http://dx.doi.org/10.1088/0034-4885/76/4/044201}{{\em Rept. Prog.
  Phys.} {\bfseries 76} (2013) 044201},
\href{http://arxiv.org/abs/1209.2710}{{\ttfamily arXiv:1209.2710 [hep-ph]}}.

\bibitem{Cyr-Racine:2013jua}
F.-Y. Cyr-Racine and K.~Sigurdson, ``{Limits on Neutrino-Neutrino Scattering in
  the Early Universe},''
  \href{http://dx.doi.org/10.1103/PhysRevD.90.123533}{{\em Phys. Rev.}
  {\bfseries D90} no.~12, (2014) 123533},
\href{http://arxiv.org/abs/1306.1536}{{\ttfamily arXiv:1306.1536
  [astro-ph.CO]}}.

\bibitem{Lancaster:2017ksf}
L.~Lancaster, F.-Y. Cyr-Racine, L.~Knox, and Z.~Pan, ``{A tale of two modes:
  Neutrino free-streaming in the early universe},''
  \href{http://dx.doi.org/10.1088/1475-7516/2017/07/033}{{\em JCAP} {\bfseries
  1707} (2017) 033},
\href{http://arxiv.org/abs/1704.06657}{{\ttfamily arXiv:1704.06657
  [astro-ph.CO]}}.

\bibitem{Park:2019ibn}
M.~Park, C.~D. Kreisch, J.~Dunkley, B.~Hadzhiyska, and F.-Y. Cyr-Racine,
  ``{$\Lambda$CDM or self-interacting neutrinos: How CMB data can tell the two
  models apart},'' \href{http://dx.doi.org/10.1103/PhysRevD.100.063524}{{\em
  Phys. Rev.} {\bfseries D100} no.~6, (2019) 063524},
\href{http://arxiv.org/abs/1904.02625}{{\ttfamily arXiv:1904.02625
  [astro-ph.CO]}}.

\bibitem{Knox:2019rjx}
L.~Knox and M.~Millea, ``{Hubble constant hunter\textquoteright{}s guide},''
  \href{http://dx.doi.org/10.1103/PhysRevD.101.043533}{{\em Phys. Rev. D}
  {\bfseries 101} no.~4, (2020) 043533},
  \href{http://arxiv.org/abs/1908.03663}{{\ttfamily arXiv:1908.03663
  [astro-ph.CO]}}.

\bibitem{Brdar:2020nbj}
V.~Brdar, M.~Lindner, S.~Vogl, and X.-J. Xu, ``{Revisiting neutrino
  self-interaction constraints from $Z$ and $\tau$ decays},''
  \href{http://dx.doi.org/10.1103/PhysRevD.101.115001}{{\em Phys. Rev.}
  {\bfseries D101} no.~11, (2020) 115001},
\href{http://arxiv.org/abs/2003.05339}{{\ttfamily arXiv:2003.05339 [hep-ph]}}.

\bibitem{Lyu:2020lps}
K.-F. Lyu, E.~Stamou, and L.-T. Wang, ``{Self-interacting neutrinos: solution
  to Hubble tension versus experimental constraints},''
\href{http://arxiv.org/abs/2004.10868}{{\ttfamily arXiv:2004.10868 [hep-ph]}}.

\bibitem{Berbig:2020wve}
M.~Berbig, S.~Jana, and A.~Trautner, ``{The Hubble tension and a renormalizable
  model of gauged neutrino self-interactions},''
  \href{http://dx.doi.org/10.1103/PhysRevD.102.115008}{{\em Phys. Rev.}
  {\bfseries D102} no.~11, (2020) 115008},
\href{http://arxiv.org/abs/2004.13039}{{\ttfamily arXiv:2004.13039 [hep-ph]}}.

\bibitem{Riess:2019cxk}
A.~G. Riess, S.~Casertano, W.~Yuan, L.~M. Macri, and D.~Scolnic, ``{Large
  Magellanic Cloud Cepheid Standards Provide a 1\% Foundation for the
  Determination of the Hubble Constant and Stronger Evidence for Physics beyond
  $\Lambda$CDM},'' \href{http://dx.doi.org/10.3847/1538-4357/ab1422}{{\em
  Astrophys. J.} {\bfseries 876} no.~1, (2019) 85},
\href{http://arxiv.org/abs/1903.07603}{{\ttfamily arXiv:1903.07603
  [astro-ph.CO]}}.

\bibitem{Aghanim:2019ame}
{\bfseries Planck} Collaboration, N.~Aghanim {\em et~al.}, ``{Planck 2018
  results. V. CMB power spectra and likelihoods},''
  \href{http://dx.doi.org/10.1051/0004-6361/201936386}{{\em Astron. Astrophys.}
  {\bfseries 641} (2020) A5},
\href{http://arxiv.org/abs/1907.12875}{{\ttfamily arXiv:1907.12875
  [astro-ph.CO]}}.

\bibitem{Alam:2016hwk}
{\bfseries BOSS} Collaboration, S.~Alam {\em et~al.}, ``{The clustering of
  galaxies in the completed SDSS-III Baryon Oscillation Spectroscopic Survey:
  cosmological analysis of the DR12 galaxy sample},''
  \href{http://dx.doi.org/10.1093/mnras/stx721}{{\em Mon. Not. Roy. Astron.
  Soc.} {\bfseries 470} no.~3, (2017) 2617--2652},
  \href{http://arxiv.org/abs/1607.03155}{{\ttfamily arXiv:1607.03155
  [astro-ph.CO]}}.

\bibitem{Ross:2014qpa}
A.~J. Ross, L.~Samushia, C.~Howlett, W.~J. Percival, A.~Burden, and M.~Manera,
  ``{The clustering of the SDSS DR7 main Galaxy sample \textendash{} I. A 4 per
  cent distance measure at $z = 0.15$}''
  \href{http://dx.doi.org/10.1093/mnras/stv154}{{\em Mon. Not. Roy. Astron.
  Soc.} {\bfseries 449} no.~1, (2015) 835--847},
  \href{http://arxiv.org/abs/1409.3242}{{\ttfamily arXiv:1409.3242
  [astro-ph.CO]}}.

\bibitem{Beutler:2011hx}
F.~Beutler, C.~Blake, M.~Colless, D.~Jones, L.~Staveley-Smith, L.~Campbell,
  Q.~Parker, W.~Saunders, and F.~Watson, ``{The 6dF Galaxy Survey: Baryon
  Acoustic Oscillations and the Local Hubble Constant},''
  \href{http://dx.doi.org/10.1111/j.1365-2966.2011.19250.x}{{\em Mon. Not. Roy.
  Astron. Soc.} {\bfseries 416} (2011) 3017--3032},
  \href{http://arxiv.org/abs/1106.3366}{{\ttfamily arXiv:1106.3366
  [astro-ph.CO]}}.

\bibitem{Scolnic:2017caz}
D.~Scolnic {\em et~al.}, ``{The Complete Light-curve Sample of
  Spectroscopically Confirmed SNe Ia from Pan-STARRS1 and Cosmological
  Constraints from the Combined Pantheon Sample},''
  \href{http://dx.doi.org/10.3847/1538-4357/aab9bb}{{\em Astrophys. J.}
  {\bfseries 859} no.~2, (2018) 101},
  \href{http://arxiv.org/abs/1710.00845}{{\ttfamily arXiv:1710.00845
  [astro-ph.CO]}}.

\bibitem{Brinckmann:2020bcn}
T.~Brinckmann, J.~H. Chang, and M.~LoVerde, ``{Self-interacting neutrinos, the
  Hubble parameter tension, and the Cosmic Microwave Background},''
  \href{http://arxiv.org/abs/2012.11830}{{\ttfamily arXiv:2012.11830
  [astro-ph.CO]}}.

\bibitem{Das:2020xke}
A.~Das and S.~Ghosh, ``{Flavor-specific Interaction Favours Strong Neutrino
  Self-coupling},''
\href{http://arxiv.org/abs/2011.12315}{{\ttfamily arXiv:2011.12315 [hep-ph]}}.

\bibitem{Mazumdar:2020ibx}
A.~Mazumdar, S.~Mohanty, and P.~Parashari, ``{Flavour specific neutrino
  self-interaction: $H_0$ tension and IceCube},''
\href{http://arxiv.org/abs/2011.13685}{{\ttfamily arXiv:2011.13685 [hep-ph]}}.

\bibitem{Aartsen:2020aqd}
{\bfseries IceCube} Collaboration, M.~Aartsen {\em et~al.}, ``{Characteristics
  of the diffuse astrophysical electron and tau neutrino flux with six years of
  IceCube high energy cascade data},''
  \href{http://dx.doi.org/10.1103/PhysRevLett.125.121104}{{\em Phys. Rev.
  Lett.} {\bfseries 125} no.~12, (2020) 121104},
  \href{http://arxiv.org/abs/2001.09520}{{\ttfamily arXiv:2001.09520
  [astro-ph.HE]}}.

\bibitem{Lewis:1999bs}
A.~Lewis, A.~Challinor, and A.~Lasenby, ``{Efficient computation of CMB
  anisotropies in closed FRW models},''
  \href{http://dx.doi.org/10.1086/309179}{{\em Astrophys. J.} {\bfseries 538}
  (2000) 473--476}, \href{http://arxiv.org/abs/astro-ph/9911177}{{\ttfamily
  arXiv:astro-ph/9911177}}.

\bibitem{Hannestad:2000gt}
S.~Hannestad and R.~J. Scherrer, ``{Selfinteracting warm dark matter},''
  \href{http://dx.doi.org/10.1103/PhysRevD.62.043522}{{\em Phys. Rev. D}
  {\bfseries 62} (2000) 043522},
  \href{http://arxiv.org/abs/astro-ph/0003046}{{\ttfamily
  arXiv:astro-ph/0003046}}.

\bibitem{Hannestad:2004qu}
S.~Hannestad, ``{Structure formation with strongly interacting neutrinos -
  Implications for the cosmological neutrino mass bound},''
  \href{http://dx.doi.org/10.1088/1475-7516/2005/02/011}{{\em JCAP} {\bfseries
  0502} (2005) 011},
\href{http://arxiv.org/abs/astro-ph/0411475}{{\ttfamily arXiv:astro-ph/0411475
  [astro-ph]}}.

\bibitem{Ma:1995ey}
C.-P. Ma and E.~Bertschinger, ``{Cosmological perturbation theory in the
  synchronous and conformal Newtonian gauges},''
  \href{http://dx.doi.org/10.1086/176550}{{\em Astrophys. J.} {\bfseries 455}
  (1995) 7--25},
\href{http://arxiv.org/abs/astro-ph/9506072}{{\ttfamily arXiv:astro-ph/9506072
  [astro-ph]}}.

\bibitem{Aghanim:2018oex}
{\bfseries Planck} Collaboration, N.~Aghanim {\em et~al.}, ``{Planck 2018
  results. VIII. Gravitational lensing},''
  \href{http://dx.doi.org/10.1051/0004-6361/201833886}{{\em Astron. Astrophys.}
  {\bfseries 641} (2020) A8}, \href{http://arxiv.org/abs/1807.06210}{{\ttfamily
  arXiv:1807.06210 [astro-ph.CO]}}.

\bibitem{Handley:2015vkr}
W.~J. Handley, M.~P. Hobson, and A.~N. Lasenby, ``{polychord: next-generation
  nested sampling},'' \href{http://dx.doi.org/10.1093/mnras/stv1911}{{\em Mon.
  Not. Roy. Astron. Soc.} {\bfseries 453} no.~4, (2015) 4385--4399},
\href{http://arxiv.org/abs/1506.00171}{{\ttfamily arXiv:1506.00171
  [astro-ph.IM]}}.

\bibitem{Handley:2015fda}
W.~Handley, M.~Hobson, and A.~Lasenby, ``{PolyChord: nested sampling for
  cosmology},'' \href{http://dx.doi.org/10.1093/mnrasl/slv047}{{\em Mon. Not.
  Roy. Astron. Soc.} {\bfseries 450} no.~1, (2015) L61--L65},
  \href{http://arxiv.org/abs/1502.01856}{{\ttfamily arXiv:1502.01856
  [astro-ph.CO]}}.

\bibitem{Lewis:2002ah}
A.~Lewis and S.~Bridle, ``{Cosmological parameters from CMB and other data: A
  Monte Carlo approach},''
  \href{http://dx.doi.org/10.1103/PhysRevD.66.103511}{{\em Phys. Rev. D}
  {\bfseries 66} (2002) 103511},
  \href{http://arxiv.org/abs/astro-ph/0205436}{{\ttfamily
  arXiv:astro-ph/0205436}}.

\bibitem{Lewis:2013hha}
A.~Lewis, ``{Efficient sampling of fast and slow cosmological parameters},''
  \href{http://dx.doi.org/10.1103/PhysRevD.87.103529}{{\em Phys. Rev. D}
  {\bfseries 87} (2013) 103529},
\href{http://arxiv.org/abs/1304.4473}{{\ttfamily arXiv:1304.4473
  [astro-ph.CO]}}.

\bibitem{Handley}
W.~Handley, ``{CosmoChord: Planck 2018 update},''
  \href{http://arxiv.org/abs/https://doi.org/10.5281/zenodo.3370086}{{\ttfamily
  https://doi.org/10.5281/zenodo.3370086}}.

\bibitem{Mead:2020vgs}
A.~Mead, S.~Brieden, T.~Tröster, and C.~Heymans, ``{HMcode-2020: Improved
  modelling of non-linear cosmological power spectra with baryonic feedback},''
\href{http://arxiv.org/abs/2009.01858}{{\ttfamily arXiv:2009.01858
  [astro-ph.CO]}}.

\bibitem{Lewis:2019xzd}
A.~Lewis, ``{GetDist: a Python package for analysing Monte Carlo samples},''
  \href{http://arxiv.org/abs/1910.13970}{{\ttfamily arXiv:1910.13970
  [astro-ph.IM]}}.
\url{https://getdist.readthedocs.io}.

\bibitem{jeffreys1998theory}
H.~Jeffreys, {\em The Theory of Probability}.
\newblock Oxford Classic Texts in the Physical Sciences. OUP Oxford, 1998.
\newblock \url{https://books.google.dk/books?id=vh9Act9rtzQC}.

\bibitem{Akaike1975}
H.~{Akaike}, ``A new look at the statistical model identification,''
  \href{http://dx.doi.org/10.1109/TAC.1974.1100705}{{\em IEEE Transactions on
  Automatic Control} {\bfseries 19} no.~6, (1974) 716--723}.

\end{thebibliography}\endgroup

\end{document}